\documentclass[useAMS,usenatbib]{mn2e}
\usepackage{graphicx}
\usepackage{epsfig}
\usepackage{epsf}
\usepackage{amsmath}
\usepackage{amssymb}
\usepackage{stfloats}
\title[SN 2009ip]{The Unprecedented 2012 Outburst of SN~2009ip: A
  Luminous Blue Variable Star Becomes a True Supernova}

\author[Mauerhan et al.]{Jon C. Mauerhan$^{1}$\thanks{E-mail:
    mauerhan@as.arizona.edu}, Nathan Smith$^{1}$, Alexei
  V. Filippenko$^{2}$, Kyle B. Blanchard$^{2}$, \newauthor Peter K.
  Blanchard$^{2}$, Chadwick F. E. Casper$^{2}$, S. Bradley
  Cenko$^{2}$, Kelsey I. Clubb$^{2}$, \newauthor Daniel P. Cohen$^{2}$,
  Kiera L. Fuller$^{2}$, Gary Z. Li$^{2}$, and Jeffrey M. Silverman$^{2,3}$ \\
  $^{1}$Steward Observatory, University of Arizona, 933 N. Cherry
  Ave., Tucson, Arizona 85721, USA \\ $^2$Department of Astronomy,
  University of California, Berkeley, CA 94720-3411, USA \\ 
  $^3$Department of Astronomy, University of Texas, Austin, 
  TX 78712-0259, USA }
\begin{document}

\pagerange{\pageref{firstpage}--\pageref{lastpage}} \pubyear{2012}

\maketitle

\label{firstpage}

\begin{abstract}

  Some reports of supernova (SN) discoveries turn out not to be true
  core-collapse explosions.  One such case was SN~2009ip, which was
  recognised to be the eruption of a luminous blue variable (LBV)
  star. This source had a massive (50--80\,M$_{\odot}$), hot progenitor
  star identified in pre-explosion data, it had documented evidence of
  pre-outburst variability, and it was subsequently discovered to have
  a second outburst in 2010.  In 2012, the source entered its third
  known outburst.  Initial spectra showed the same narrow-line
  profiles as before, suggesting another LBV-like eruption.  We
  present new photometry and spectroscopy of SN~2009ip, indicating
  that the 2012 outburst transitioned into a genuine SN explosion. The
  most striking aspect of these data is that unlike any previous
  episodes, the spectrum developed Balmer lines with very broad
  P-Cygni profiles characteristic of normal Type II supernovae
  (SNe~II), in addition to overlying narrow emission components. The
  emission lines exhibit unprecedented (for any known nonterminal 
  LBV-like eruption) full width at half-maximum
  intensity values of $\sim 8000$\,km\,s$^{-1}$, while the absorption
  components seen just before the main brightening had blue wings
  extending out to $-$13,000\,km\,s$^{-1}$.  These velocities are
  typical of core-collapse SN explosions, but have never been associated
  with emission lines from a nonterminal LBV-like eruption.  
  SN~2009ip is the first object to
  have both a known massive blue progenitor star and LBV-like
  eruptions with accompanying spectra observed a few years prior to
  becoming a SN.  Immediately after the broad lines first appeared,
  the peak absolute magnitude of $M_V \approx -14.5$ was fainter than
  that of normal SNe~II.  However, after a brief period of fading, the 
  source quickly brightened again to $M_R=-17.5$ mag in $\sim 2$ days,
  suggesting a causal link to the prior emergence of the broad-line
  spectrum.  Once the bright phase began, the broad lines mostly
  disappeared, and the spectrum resembled the early optically thick
  phases of luminous SNe~IIn. The source reached a peak brightness of
  $-18$ mag about 2 weeks later, after which broad emission lines
  again developed in the spectrum as the source faded.  We conclude
  that the most recent 2012 outburst of SN~2009ip was the result of a
  true core-collapse SN~IIn that occurred when the progenitor star was
  in an LBV-like outburst phase, and where the SN was initially faint
  and then rapidly brightened due to interaction with circumstellar
  material.  The pulsational pair instability, LBV-like eruptions, or
  other instabilities due to late nuclear burning phases in massive
  stars may have caused the multiple pre-SN eruptions.
  \end{abstract}

\begin{keywords}
  circumstellar matter --- stars: evolution --- stars: winds, outflows
  --- supernovae: general --- supernovae: individual (SN~2009ip)
\end{keywords}

\section{Introduction}

Some very massive stars undergo sporadic luminous outbursts
accompanied by episodic ejection of matter during their post-main-sequence
evolution, in a few cases immediately before the final supernova (SN)
explosion (e.g., Smith et al. 2008a, 2010a, 2010b; Smith \& Owocki 2006;
Gal-Yam \& Leonard 2009). Strong evidence for this has been provided
by observations of interacting supernovae (Type IIn/Ibn SNe), the
spectra of which exhibit relatively narrow (narrow and ``intermediate
width'') emission lines that are generated by the interaction between
the SN blast wave and dense circumstellar material (CSM) ejected by
the stellar progenitor in the years to decades before the final
explosion. Depending on the nature of the star, the CSM may be created
via strong superwinds that last for millennia, in the case of extreme
red supergiant progenitors (Smith et al.\ 2009), or by discrete
outbursts that occur rapidly on month-to-year timescales (see Chugai
et al.\ 2004; Chugai \& Danziger 1994; Smith et al.\ 2010a,
2008a). The episodic outbursts are reminiscent of luminous
blue variable (LBV) stars. 

Several such events have been observed directly, including the historic eruption of 
$\eta$ Carinae in our own Galaxy during the mid-19th century (Smith \&
Frew 2011), as well as the more recently discovered extragalactic
analogs, the so-called ``SN impostors'' (Van Dyk et al.\ 2000; Smith
et al.\ 2011). These eruptive/explosive events can collectively
liberate significantly more material than the total integrated mass
lost via steady line-driven winds during the lifetime of a massive
star (Smith \& Owocki 2006), although the physical mechanism or
trigger that causes the eruptions remains uncertain. The physics of this
explosive mode of mass loss is poorly understood, and almost
completely unaccounted for in current stellar evolution models.

Eruptive pre-SN mass loss from SN progenitors can be probed indirectly by
observations of interacting SNe. In the extraordinary case of the Type
Ibn SN~2006jc, a pre-SN outburst from the progenitor was observed
directly, two years before core collapse (Pastorello et al.\
2007). The precursor outburst, observed only photometrically, was
believed to be the source of dense CSM emission and X-rays detected
throughout the evolution of the SN spectrum (Pastorello et al.\ 2007;
Foley et al.\ 2007; Smith et al.\ 2008b; Immler et al. 2008). Owing to
the H-poor abundances of the CSM, the progenitor of SN~2006jc was
believed to be a Wolf-Rayet (WR) star. Since WRs are not typically
associated with luminous outbursts, the eruptive progenitor was suggested to
have recently transitioned from the LBV phase into a WR (Foley et al.\
2007).  While LBVs provide the only known precedent for the eruptive
pre-SN mass loss needed to make luminous SNe~IIn (e.g., Smith et al.\
2010a, 2008a), this suggestion has been controversial because it
directly contradicts current expectations of standard stellar evolution models
(Heger et al.\ 2003; Langer et al.\ 1994; Maeder \& Meynet 2000).

\begin{figure}
\includegraphics[width=3.3in]{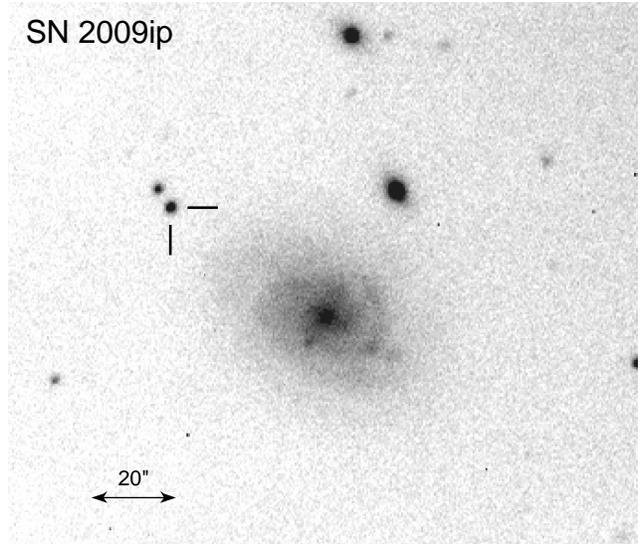}
\caption{$R$-band Kuiper/MONT4K image of SN~2009ip and its host galaxy
  NGC 7259 on 2012 Sep. 24. North is up and east is to the left.}
\label{fig:im1}
\end{figure}

The SN impostor SN~2009ip, shown in Figure~\ref{fig:im1}, provides a
rare case where \textit{multiple} outbursts were observed
approximately one year apart (Smith et al.\ 2010b; Drake et al.\
2010). Archival images from the \textit{Hubble Space Telescope (HST)}
revealed a luminous blue progenitor at the location of the transient,
having photometry consistent with a stellar luminosity of
log$(L/{\rm L}_{\odot}) \approx 5.9$ and an initial mass of
50--80\,M$_{\odot}$, indicating that the source is likely to be an LBV
(Smith et al.\ 2010b).  Subsequently, Foley et al.\ (2011) provided
their own analysis of the same {\it HST} data and found a consistent
result, indicating a progenitor with an initial mass above
60\,M$_{\odot}$. The first 2009 outburst was very brief compared to
other LBV-like eruptions, lasting only a few days (Smith et al.\
2010b), instead of a few months or more like most LBV eruptions
(Smith et al.\ 2011).  While narrow Lorentzian emission-line profiles
indicated that most of the ejected mass was expanding at around
600\,km\,s$^{-1}$, higher speeds \textit{seen only in absorption}
suggested a small amount of ejecta moving at speeds up to
$\sim 5000$\,km\,s$^{-1}$ or more (Smith et al.\ 2010b; Foley et al.\
2011).  This provided an interesting likeness to $\eta$~Car, which
also had a small fraction of its ejected mass moving at similarly high
speeds (Smith 2008, 2012).

On 2012 July 24 (UT dates are used throughout this paper), SN~2009ip
was discovered entering its third known outburst. The first calibrated
photometric measurements on 2012 Aug. 14 revealed the source to have
a brightness of $M_V \approx -14.5$ mag (Drake et al.\ 2012). The earliest
spectrum of the 2012 outburst was obtained on 2012 Aug. 24 (Foley et
al.\ 2012); it exhibited narrow Balmer emission features similar to
those observed during the previous outbursts, having Lorentzian
profiles with full width at half-maximum intensity (FWHM) $\approx$
640\,km\,s$^{-1}$.

\begin{table}\begin{center}\begin{minipage}[bp]{3.2in} \setlength{\tabcolsep}{4.8pt}
      \caption{Photometry of SN~2009ip. Uncertainties represent
        the standard deviation of the zero-point magnitudes derived
        using APASS photometry of 4--5 field stars. (a) Photometry from
        Brimacombe (2012). (b) Kuiper/MONT4K. (c) KAIT. All others: Lick/Nickel. All dates are UT.} %}
\centering
  \begin{tabular}[bp]{@{}lcccc@{}} 
  \hline
   Date      & $B$  & $R$ & $I$   \\                            
    (JD$-$2,450,000)       & (mag) & (mag) & (mag) \\
\hline
\hline
Aug. 28 (6167.83)          &  $17.38~(0.11)$  &  $16.89~(0.09)$  & $17.09~(0.12)$ \\
Sep. 05 (6175.81)          &  $17.20~(0.14)$  &$16.73~(0.22)$  & $16.85~(0.11)$ \\
Sep. 08 (6178.81)          &  $17.15~(0.15)$ &$16.71~(0.19)$  & $17.06~(0.10)$ \\
Sep. 10 (6180.79)          &  $17.22~(0.15)$ & $16.74~(0.19)$  & $16.92~(0.25)$ \\
Sep. 18 (6188.78)          &  $18.19~(0.17)$ & $17.46~(0.20)$  & $17.60~(0.20)$ \\
Sep. 20 (6190.81)$^c$ &         :::                  & $17.31~(0.24)$  & $17.53~(0.20)$  \\
Sep. 21 (6191.76)          &  $18.02~(0.21)$ &$17.50~(0.24)$  & $17.79~(0.20)$  \\
Sep. 24 (6194.10)$^a$ &        :::                  &        :::                 & $18.20~(0.25)$ \\
Sep. 24 (6194.75)$^c$ &         :::                 &$17.36~(0.18)$   & $17.44~(0.22)$  \\
Sep. 24 (6194.69)$^b$ &  $17.84~(0.12)$ &$17.35~(0.14)$    &           :::               \\
Sep. 25 (6195.07)$^a$ &        :::                  &        :::                 & $16.60~(0.21)$ \\
Sep. 25 (6195.70)$^a$ &        :::                  &        :::                 & $15.00~(0.21)$ \\
Sep. 25 (6195.81)$^c$ &         :::                  &$15.18~(0.15)$& $15.11~(0.22)$  \\
Sep. 26 (6196.02)$^a$ &        :::                  &         :::                 & $15.60~(0.19)$ \\
Sep. 26 (6196.76)          &        :::                  & $14.57~(0.14)$  & $14.49~(0.15)$ \\
Sep. 29 (6199.76)          &  $14.20~(0.08)$   & $ 14.02~(0.09)$ & $13.91~(0.12)$  \\
Sep. 30 (6200.60)$^c$ &  :::                            &  $ 13.96~(0.14)$ & $13.77~(0.14)$ \\  
Oct. 01 (6201.60)$^c$  &  :::                            &  $  13.88~(0.13)$ & $13.74~(0.13)$ \\ 
Oct. 03 (6203.71)          &   $13.87~(0.09)$ & $13.73~(0.10)$  &  $13.59~(0.10)$  \\
Oct. 07 (6207.72)          &  $13.82~(0.10) $  &$13.64~(0.10)$   & $13.44~(0.11)$ \\
Oct. 09 (6209.55)$^c$          &  :::                             & $13.69~(0.11) $ & ::: \\
Oct. 11 (6211.72)      &  $13.86~(0.14) $      &$13.72~(0.13)$   & $13.51~(0.15)$ \\
Oct. 14 (6214.60)$^c$ & $14.31~(0.15)$ & $13.71~(0.15)$ & ::: \\
Oct. 18 (6218.60)$^c$ & :::                         & $ 13.92~(0.14)$ & ::: \\
Oct. 19 (6219.69)      & $14.38~(0.13)$         &$13.98~(0.14)$ & $13.67~(0.15)$ \\
Oct. 24 (6224.73)         & $14.71~(0.15)$ & $14.21~(0.14)$ & $13.97~(0.15)$ \\
Oct. 30 (6230.69)  & $15.46~(0.16)$ & $14.80~(0.15)$ & $14.64~(0.16)$ \\
Nov. 03 (6234.70) & $ 15.16~(0.13)$ & $14.61~(0.13)$  & $14.41~(0.15)$ \\
Nov. 04 (6235.64) & $15.14~(0.14)$ & $14.58~(0.13)$   & $14.40~(0.14)$ \\
Nov. 05 (6236.63) & $15.17~(0.12)$& $14.60~(0.14)$  & $14.40~(0.14)$ \\
Nov. 08 (6239.65) & $15.52~(0.11) $& $14.85~(0.12)$ & $14.66~(0.11)$ \\
Nov. 14 (6245.60) & $16.22~(0.14)$ &$15.32~(0.12) $& $15.15~(0.13)$ \\
Nov. 25 (6256.59) & $16.91~(0.17)$ & $15.71~(0.13)$ & $15.48~(0.15)$ \\                                                            
Nov. 28 (6259.65) & $16.98~(0.21)$ & $15.82~(0.18)$ & $15.55~(0.19)$ \\                          
Dec. 07 (6268.60)  & $17.77~(0.20)$ & $16.10~(0.16)$ & $15.83~(0.17)$ \\   
Dec. 21 (6282.57) & $19.07~(0.29)$ & $17.33~(0.24)$ & $17.39~(0.22)$ \\
\hline
\end{tabular} \end{minipage} \end{center}
\end{table}

Here, we provide spectroscopic evidence that the latest outburst has
actually developed extremely broad emission lines, consistent with
those of a true core-collapse SN (Smith \& Mauerhan 2012a), in
addition to the narrow lines.  Specifically, we report the emergence
of strong P-Cygni profiles having broad emission components with FWHM
$\approx$ 8000\,km\,s$^{-1}$ and absorption-wing velocities of
$\sim$ 13,000\,km\,s$^{-1}$ and higher in some cases.  The large widths
measured for the broad emission components, which likely represent the
bulk velocity of the outflow, are unprecedented for any known
nonterminal LBV-like eruption, and were not seen in previous spectra
of this same object. Although some broad absorption was visible during
previous outbursts, this is the first time that broad
\textit{emission} has been detected, and it marks a significant and
new qualitative change in the object.

While the new spectra of SN~2009ip look like those of a true
core-collapse SN, the absolute magnitudes that were initially reported
were less luminous than those of normal SNe, leading some to conclude
prematurely that it is not a genuine SN (Margutti et al.\ 2012a;
Martin et al. 2012).  The initial peak absolute magnitude of
$-$14.5 measured in the first week of September was, however, more
luminous than that of some of the faintest known examples of
core-collapse SNe~II-P such as SN~1999br (Pastorello et al. 2004). The
faintness and slow rise at early times could resemble some well-known
SNe from blue supergiant progenitors that are initially faint,
including SN~1987A (e.g., Arnett 1989, and references therein).
The conclusion that it is not a core-collapse event appears to have
been unjustified, because a few days later SN~2009ip began to brighten
rapidly, reaching a luminosity consistent with a true SN after all
(Brimacombe 2012; Margutti et al. 2012b; Smith \& Mauerhan 2012b).  In
this paper, we present both new spectra and new photometry of the most
recent outburst of SN~2009ip, and we discuss the interesting
implications of this unprecedented event.

\section{Observations and Results}
\subsection{Photometry}

\begin{figure}
\includegraphics[width=3.3in]{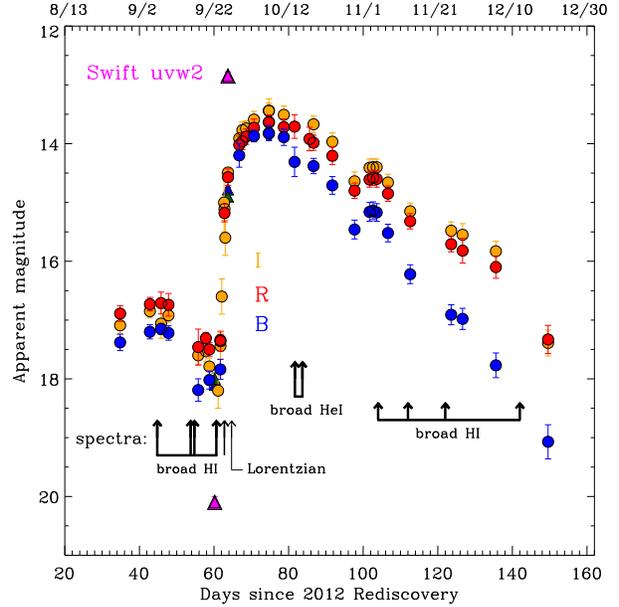}
\caption{Light curves of SN~2009ip, including our measurements from
  Table~1, and reported data from Brimacombe
  (2012) and Margutti et al. (2012b). The data are plotted with
  respect to Julian Day 2,456,133 (the discovery date of the most
  recent outburst). Epochs with accompanying spectra (presented in
  \S2.2) are marked by arrows. The upper abscissa is labeled with numeric 
  month/day dates.}
\label{fig:lca}
\end{figure}

\begin{figure*}
\includegraphics[width=6.3in]{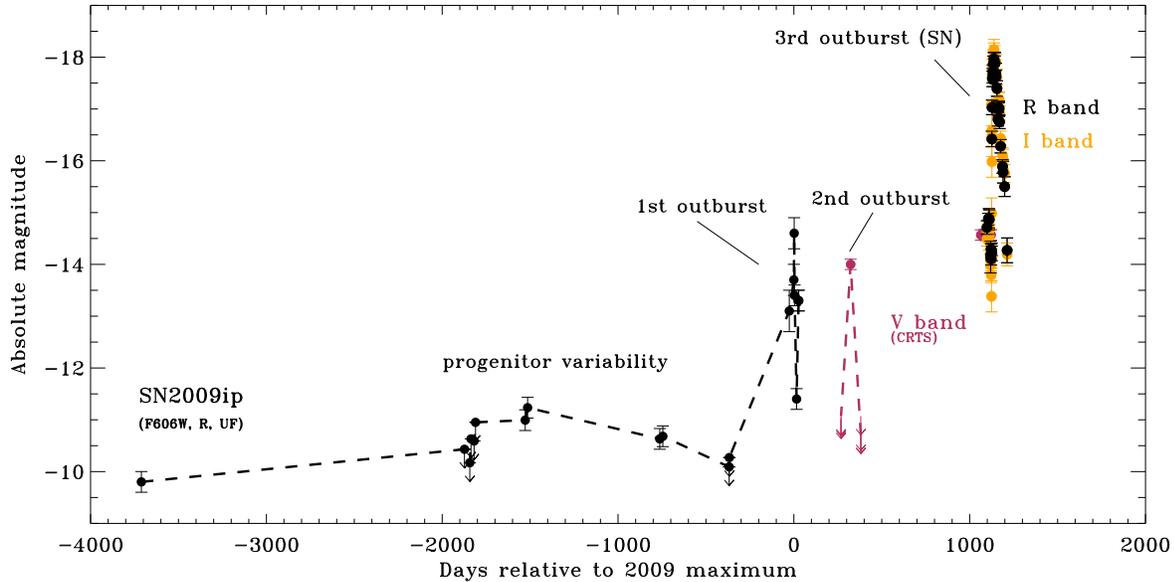}
\caption{Absolute magnitude light curve of SN~2009ip, including
  archival {\it HST} data, and the ground-based $V$, $R$, $I$, and
  unfiltered photometry (see Smith et al. 2010b).}
\label{fig:lc}
\end{figure*}

We began photometrically monitoring the 2012 outburst of SN~2009ip on
Aug. 28 using the 1\,m Nickel telescope and CCD imager at Lick
Observatory.  Measurements at 23 epochs through 2012 Dec.~21 were
obtained with $B$, $R$, and $I$ filters. Images of SN~2009ip in the
$B$ and $R$ bands were also obtained on 2012 Sep. 24 using the
MONT4K imager on the Kuiper 61\,in telescope on Mt. Bigelow in Tucson,
Arizona. In addition, 8 epochs of $B$, $R$, and $I$ photometry were
acquired between 2012 Sep. 20 and Oct. 18, using the 0.76~m
Katzman Automatic Imaging Telescope (KAIT) at Lick Observatory
(Filippenko et al. 2001). 

Photometric measurements were extracted using standard
aperture-photometry techniques and were calibrated using photometry of
4--5 field stars in the same image as the SN. For calibration
purposes, we used magnitudes from the American Association of Variable
Star Observers (AAVSO) Photometric All-Sky Survey
(APASS\footnote{http://www.aavso.org/apass}). Initially, we had used
photometry from the USNO-B1 catalog for calibration, which led to
imprecise results, especially for the $B$ band. We subsequently became
aware of the availability and superior quality of APASS photometry
(Brimacombe 2012; Prieto et al. 2012), and thus used this for our
calibration.  We also performed point-spread function (PSF) fitting
photometry of SN~2009ip and surrounding fields stars using the
IDL-based code Starfinder. The results from both techniques are fully
consistent, well within the uncertainties. 

Table 1 lists our Lick/Nickel, Kuiper/MONT4K, and KAIT photometry of
SN~2009ip, and also includes $I$-band measurements reported by
Brimacombe (2012). Our $R$ and $I$ photometry was derived by
converting the $i$ and $r$ APASS values (Sloan system) to the Johnson
system, following Lupton
(2005)\footnote{http://www.sdss.org/dr5/algorithms/sdssUBVRITransform.html}.
Absolute magnitudes were derived adopting a distance modulus of 31.55
mag for the host galaxy NGC 725 and an extinction value of $A_R =
0.051$ mag (Smith et al. 2010b), which was converted to $A_B$ and
$A_I$ following Cardelli, Clayton, \& Mathis (1989).

The new light curve of SN~2009ip is shown in Figure~\ref{fig:lca},
which also contains several measurements from recent reports,
including space-based UV photometry from \textit{Swift} in the
\textit{uvw2} band (Margutti et al. 2012a,b) having a central
wavelength of 1928\,\AA.  The complete absolute magnitude light curve
is shown in Figure~\ref{fig:lc}, which also includes the photometry
originally presented by Smith et al. (2010b), and measurements of the
second 2010 outburst from the Catalina Real-time Transient Survey
(CRTS\footnote{http://nesssi.cacr.caltech.edu/catalina/current.html.})
0.5\,m Schmidt telescope, which is part of the Siding Spring Survey
(SSS).

The early photometry revealed a luminous and highly variable
progenitor star at the location of the transient (Smith et al.\ 2010b)
during the years leading up to the first observed outburst in 2009.
We recently noticed that the progenitor star was detected in
photographic plates from the second epoch of the southern plate survey
(USNO-A2, not shown here) and appears to have been active on 1996
Aug. 14, which is earlier than previous studies have recognised. At
this early stage the progenitor appears to have a very approximate
brightness of $R \approx 18$--19 mag ($-12 < M_R < -11$ mag),
comparable to magnitudes reached during the pre-outburst variability
phase marked in Figure~\ref{fig:lc}.

During the first outburst in 2009, the source brightened to an
absolute magnitude of $\sim -14.5$ mag and faded on a relatively rapid
timescale of several days (Smith et al.\ 2010b). After a quiescent
period of $\sim 1$ year, a second outburst occurred, achieving a
comparable luminosity and fading on a similarly rapid
timescale. Nearly 2 years later, the latest outburst occurred, first
detected on 2012 July 24, although calibrated photometry was
apparently not publicly available at that time. Subsequent photometry
on 2012 Aug. 14 revealed an absolute $V$ magnitude of $\sim
-14.5$. Approximately one month later, SN~2009ip appeared to fade by
$\sim 1$ mag over a time span of 10 days
(Figure~\ref{fig:lca}). However, on Sep. 25, a large ($\sim 3$--4
mag) increase in brightness began to develop rapidly in the optical
bands, accompanied by an even bigger ($\sim 7$ mag) increase in the UV
bands (Brimacombe 2012; Margutti et al. 2012a,b).  By 2012 Oct. 7,
SN~2009ip had brightened and reached a plateau at $M_R \approx -18$
mag, which implies a peak luminosity of $\sim 1.3 \times 10^{9}\, {\rm
  L}_{\odot}$.  The source subsequently began to decline and became
gradually redder in color. However, beginning Nov.~3 there was a
small temporary brightness increase lasting for several days before
the decline resumed and continued through Dec.~21, with a slight but apparent
change in the decline rate between Nov.~25 and Dec.~21. 

\subsection{Spectroscopic Observations}

Spectra of SN~2009ip were obtained at multiple facilities throughout
the 2012 outburst. On Sep. 16, 17, 27, and Dec. 14, spectra were
acquired with the 2.3\,m Bok Telescope on Kitt Peak with the Boller \&
Chivens (B\&C) Spectrograph. We utilised the 400\,l\,mm$^{-1}$ and
1200\,l\,mm$^{-1}$ gratings, which provided respective spectroscopic
resolutions ($R = \lambda / \Delta\lambda$) of $\sim 1100$ and $\sim
3600$ through a 1\farcs5 wide slit. The lower-resolution spectrum was
obtained from four separate integrations of 900\,s each, and the
higher-resolution spectrum from six exposures of 1200\,s each. The low
declination of SN~2009ip (near $-29^{\circ}$) limited our observations
to be performed through a high airmass of ${\rm sec}\,z =
2.1$--2.6. Observing with the slit at the parallactic angle
(Filippenko 1982) was thus critical to avoid relative loss of blue
light owing to atmospheric dispersion. Flat-field and wavelength
calibration were performed using spectra of continuum and He-Ne-Ar
emission sources internal to the instruments. Flux calibration was
provided by observations of the A0V standard star HR~7596, with the
difference in airmass between calibration and science observations
taken into account. Data reduction and calibration were performed
using standard IRAF\footnote{IRAF: The Image Reduction and Analysis
  Facility is distributed by the National Optical Astronomy
  Observatory, which is operated by the Association of Universities
  for Research in Astronomy (AURA) under cooperative agreement with
  the National Science Foundation (NSF).} routines. For all spectra
presented in this paper, the wavelength scale was corrected for the
redshift of the host galaxy NGC 7259 ($z = 0.005715$).

\begin{figure}
\includegraphics[width=3.3in]{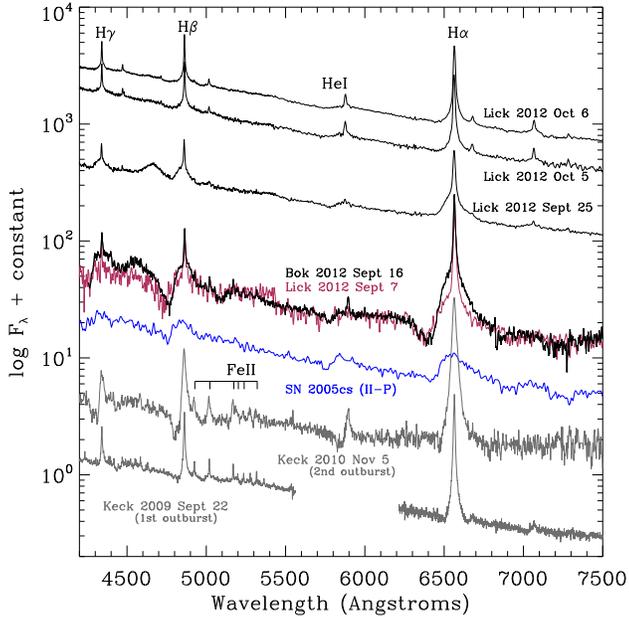}
\caption{Bok/B\&C and Lick/Kast low-resolution spectra of
  SN~2009ip. Broad lines distinguish the latest 2012 outburst (upper
  black and maroon spectra) from the prior 2009 (light grey; Smith et
  al. 2010b) and 2010 (dark grey) outbursts. However, the broad
  features diminished by Sep. 25 (top, black spectrum).  A spectrum of
  SN~2005cs (blue; Pastorello et al. 2006) is included to illustrate that the spectral
  morphology of the broad emission components of SN~2009ip is similar
  to those of a SN~II-P.}
\label{fig:spec1}
\end{figure}

\begin{figure}
\includegraphics[width=3.3in]{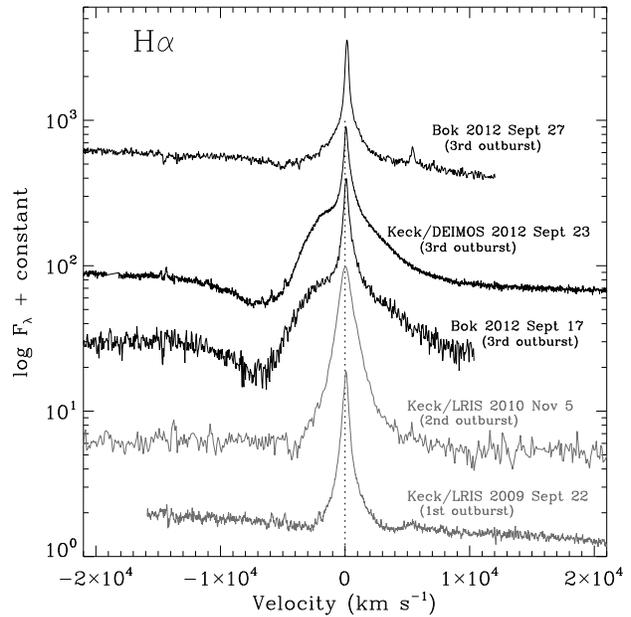}
\caption{Moderate-resolution spectra of the H$\alpha$ profile of
  SN~2009ip, plotted in velocity coordinates. On Sep. 17 and 23 the
  profile is a superposition of broad P-Cygni (FWHM $\approx
  8000$\,km\,s$^{-1}$) and narrow emission lines. The blue edge of the
  absorption trough extends to about $-$13,000\,km\,s$^{-1}$.  By
  Sep. 27 the broad components have mostly diminished, leaving 
  Lorentzian profiles. Similarly broad lines were not seen in our
  Keck/LRIS spectra of the prior two outbursts in 2009 and 2010
  (lower, grey spectra).}
\label{fig:spec2}
\end{figure}

Spectra were also obtained on 2012 Sep. 23 using the Keck-II 10\,m
telescope and the Deep Imaging Multi-Object Spectrograph (DEIMOS;
Faber et al.\ 2003). We utilised the 1200\,l\,mm$^{-1}$ grating and the
0\farcs8 slit, which provided a spectral resolution of $R \approx
4000$. Flat-field and wavelength calibration were performed using
internal continuum and arc lamps, and flux calibration was derived
from spectral measurements of the standard star BD$+$17$^\circ$4708.

Low-resolution CCD spectra of SN~2009ip were obtained on 2012 Sep. 7
and 25, Oct. 5 and 6, and Nov. 6 and 14, using the Kast spectrograph
(Miller \& Stone 1993) on the 3\,m Shane reflector at Lick
Observatory.  For our typical setup, we observed with a 300/7500
grating on the red side, a 600/4310 grism on the blue side, a D55
dichroic (giving a crossover wavelength $\sim 5500$\,\AA), and a $2''$
wide slit. The typical wavelength coverage is 3300--10,400~\AA\ with
resolutions of $\sim 11$ and $\sim 6$\,\AA\ on the red and blue sides,
respectively.

Additional low-resolution spectra were obtained on 2012 Oct. 16 and
Nov.~23 with the Bluechannel spectrograph on the Multiple Mirror
Telescope (MMT), utilising the 300\,l\,mm$^{-1}$ grating and a
1\farcs5 slit.

We also include some previously unpublished older spectra of
SN~2009ip, obtained during or between previous eruptions.  Spectra of
the prior 2010 outburst were acquired on 2010 Nov. 5 using the
Low-resolution Imaging Spectrograph (LRIS; Oke et al. 1995) on the
Keck-1 10\,m telescope, and a spectrum of SN~2009ip was obtained in
between the 2010 and 2012 events on 2011 June 26, also using MMT and
the Bluechannel spectrograph. The data reduction for these
observations followed standard techniques as described previously by
Silverman et al.\ (2012). Only the Lick/Kast spectra have been
corrected for telluric absorption.

\subsection{Overview of the Spectra}

\begin{figure}
\includegraphics[width=3.3in]{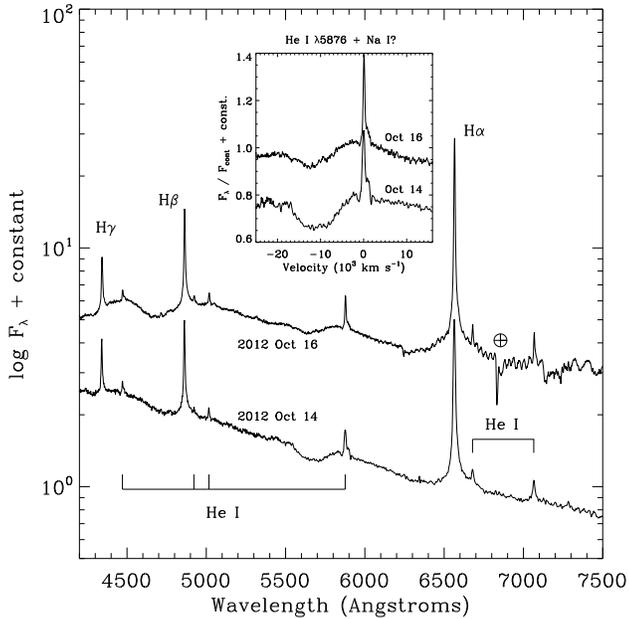}
\caption{Low-resolution MMT/Bluechannel (upper) and Lick/Kast (lower)
  spectra of SN~2009ip obtained on 2012 Oct. 16 and 14,
  respectively. Broad He~{\sc i} P-Cygni profiles have appeared and
  are accompanied by strong narrow components as well. The inset shows
  the spectrum of the He~{\sc i} $\lambda$5876 feature in
  velocity coordinates, which exhibits FWHM $\approx
  8000$\,km\,s$^{-1}$ for the main broad emission profile and a blue
  absorption edge out to $\sim$ $-$18,000\,km\,s$^{-1}$. }
\label{fig:spec_he}
\end{figure}

\begin{figure}
\includegraphics[width=3.3in]{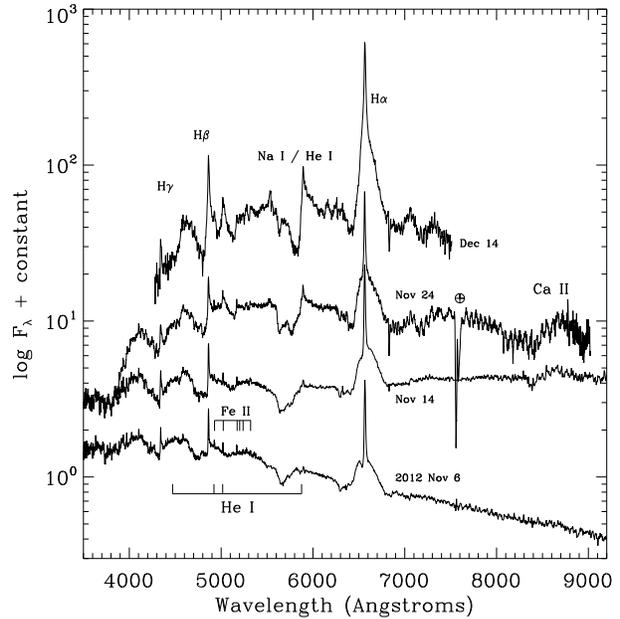}
\caption{Low-resolution spectra of SN~2009ip obtained with Lick/Kast
  (lower two spectra), MMT/Bluechannel (middle), and Bok/B\&C (top) on
  2012 Nov.~6, 14, 24, and Dec. 14, respectively. Broad H~{\sc i}
  P-Cygni profiles reappear at this stage, while the He~{\sc i} has
  mostly diminished. A strong double-absorption feature is visible
  near 5650~{\AA} on Nov.~24 and Dec.~24; it has potential
  contributions from both Na~{\sc i} and He~{\sc i}. Ca~{\sc ii} near
  8600~{\AA} emerges as a broad P-Cygni profile on Nov.~14--24.  Note
  that a P-Cygni profile for the narrow H$\alpha$ component is present
  only on Nov.~6, and it exhibits an absorption minimum velocity of
  $\sim -1300$\,km\,s$^{-1}$.}
\label{fig:spec_licklate}
\end{figure}

The spectra of SN~2009ip during the 2012 event are presented in
Figures~\ref{fig:spec1}--\ref{fig:spec_licklate}.  On Sep. 16 and 17,
the source is dominated by hydrogen Balmer emission lines that exhibit
a combination of (a) strong intermediate-width components having FWHM
velocities of $\sim 550$\,km\,s$^{-1}$ plus (b) broad P-Cygni features
having FWHM $\approx$ 8000\,km\,s$^{-1}$ and blue absorption edges
that indicate very high expansion speeds of up to
13,000\,km\,s$^{-1}$. Weak He~{\sc i} $\lambda$5876 also appears to be
detected, exhibiting the same combination of intermediate-width
emission and a broad, but weak, P-Cygni profile; however, Na~I might
also contribute to this feature.  Weak emission lines of Fe~{\sc ii}
also appear redward of H$\beta$.

The intermediate-width emission-line component during the most recent
outburst appears similar in morphology to those observed during prior
outbursts of SN~2009ip (Smith et al.\ 2010b; Foley et al.\ 2011). On
2009 Sep. 22, this component exhibited FWHM $\approx$
550\,km\,s$^{-1}$ (Smith et al. 2010b), and on 2010 Nov. 5 it had FWHM
$\approx$ 1200\,km\,s$^{-1}$. However, the underlying broad P-Cygni
profiles during the latest outburst have not been observed at any time
during the prior outbursts, and have also never before been observed
from any other purported LBV eruption (Smith et al.\ 2011).  Although
evidence for material as fast as 5000\,km\,s$^{-1}$ was seen in the
absorption components of the P-Cygni profiles of SN~2009ip's prior
eruption in 2009 (Smith et al.\ 2010b; Foley et al.\ 2011), velocities
this high have never been seen in the main emission
components. Instead, the large velocity widths of the emission lines
observed from SN~2009ip in 2012 appear very similar to those of
SNe~II-P, such as SN~2005cs (Pastorello et al. 2006), which is shown in Figure~\ref{fig:spec1} 
[data obtained from Weizmann Interactive Supernova Data Repository (Yaron \& Gal-Yam 2012)]

By Sep. 25, the broad-line components began to weaken as the continuum
strengthened.  The slope of the continuum at wavelengths between 5500
and 7500\,\AA\ appears more or less consistent with an effective
temperature of $\sim 10^4$\,K, and this is also true for the spectra
on Sep. 7--17. By Sep. 27, as the source had experienced a large
increase in brightness during which the broad-line components mostly
disappeared, the spectrum became dominated by broad-winged Lorentzian
profiles of Balmer emission, more closely resembling the spectra of
normal SNe~IIn.  The spectrum from 2 days earlier on 2012 Sep. 25
appears to have a morphology that is intermediate between the previous
broad-line spectrum from Sep. 16 and the Lorentzian spectrum on
Sep. 27.

The spectrum obtained on 2012 Oct. 16, after the source had peaked and
began to decline, reveals another dramatic change in morphology, shown
in Figure~\ref{fig:spec_he}. He~{\sc i} emission becomes very strong
at this stage, exhibiting a superposition of very broad P-Cygni
emission features and narrow lines.  The FWHM of the main broad
He~{\sc i} $\lambda$5876 emission components is $\sim
8000$\,km\,s$^{-1}$, the same as the earlier Balmer profiles. However,
the blue edge of the He~{\sc i} $\lambda$5876 absorption component
appears to extend to even higher velocities, perhaps as high as
$-$20,000\,km\,s$^{-1}$, although we must consider the possibility
that Na~{\sc i} is contributing to the absorption and/or emission
components of this feature as well. The narrow component of the
He~{\sc i} $\lambda$5876 line exhibits a P-Cygni profile, with an
absorption minimum at a velocity of $-1000$\,km\,s$^{-1}$.

\begin{figure}
\includegraphics[width=3.3in]{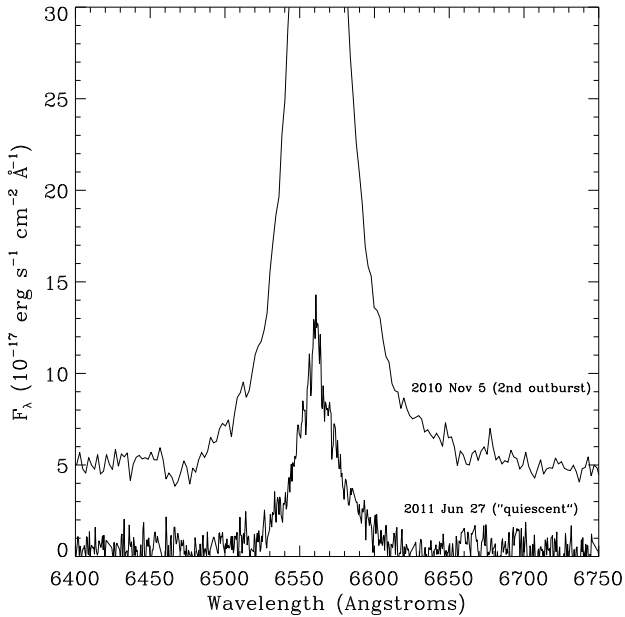}
\caption{Spectrum of H$\alpha$ obtained on 2011 June 27 (lower),
  during a relatively ``quiescent" phase in between the 2010 and 2012
  outbursts; no continuum emission was detected at this time. Although
  the line emission is much weaker than that of the prior 2010
  outburst (upper), the overall Lorentzian morphology is similar.}
\label{fig:spec_quiescent}
\end{figure}

By Nov.~6, and continuing through Dec.~14, our Lick, MMT, and Bok
spectra (Figure~\ref{fig:spec_licklate}) once again show the broad
H$\alpha$ P-Cygni emission that we observed on Sep. 7--17, and which
had disappeared in the interim. The width of the broad components
appears slightly smaller than before, having FWHM $\approx
5000$--6000\,km\,s$^{-1}$.  On Nov.~6, however, the narrow component
of the H$\alpha$ emission exhibits a P-Cygni profile as well, having
an absorption minimum near $\sim -1300$\,km\,s$^{-1}$; it diminished
in later spectra, however. By Nov.~14, broad P-Cygni emission from the
Ca~{\sc ii} triplet begins to emerge near 8600~{\AA} and strengthens
substantially by Nov.~24, which is our most recent spectrum.  Although
fringing and telluric absorption in the spectrum from Nov.~24 are
significant at wavelengths redward of H$\alpha$, the detection of
Ca~{\sc ii} is nonetheless secure.  Also appearing on Nov.~24 is a
deep double-absorption profile at the location of Na~{\sc i}/He~{\sc
  i}, in the range 5600--5800~{\AA}. The redder component was weakly
detected earlier, on Nov.~6 and 14, but becomes the deepest absorption
feature by Nov.~24, and continues to grow in strength through our
latest spectrum on Dec.~14.  It is difficult to ascertain confidently
whether this redder absorption component is the result of Na~{\sc i}
or He~{\sc i}. However, the fact that the broad He~{\sc i} features
that were seen previously (Figure~\ref{fig:spec_he}) have mostly
diminished suggests that this feature is likely dominated by Na~{\sc
  i}. The strengthening of broad Ca~{\sc ii} P-Cygni emission at the
same time might also support this interpretation, as both Ca~{\sc ii}
and Na~{\sc i} are metal lines that typically signify inner-ejecta
emission.

Finally, Figure~\ref{fig:spec_quiescent} shows a much earlier spectrum
of SN~2009ip obtained on 2011 June\ 26, between the 2010 and 2012
outbursts.  Although a factor of $\sim 100$ weaker in flux, the
morphology of the line appears similar to that of the 2010 outburst
from $\sim 7$ months earlier (also shown). No continuum flux was
detected in the 2011 spectrum, only intermediate-width H$\alpha$
emission. Although we refer to the 2011 spectrum as ``quiescent"
because the source was significantly fainter, we are not suggesting
that the progenitor was completely inactive at this time.

\section{Discussion}

\subsection{The Remarkable Sequence of Events in 2012}

The earliest reports on the spectrum of SN~2009ip for the 2012
outburst (Foley et al.\ 2012) stated that it was apparently consistent
with previous LBV-like outbursts of the same object (Smith et
al.\ 2010b; Foley et al.\ 2011).  However, this situation soon
changed.  A few weeks later, we observed the emergence of the
broad-line spectra that signaled the onset of the SN explosion (Smith
\& Mauerhan 2012a).  The broad emission components are quite
significant --- the FWHM of 8000\,km\,s$^{-1}$ for the broad emission
lines implies that a substantial fraction of the outflowing mass has
achieved these high velocities, which is beyond the capabilities of
any nonterminal super-Eddington wind-driven mechanism (Owocki et
al.\ 2004). No other known SN impostor has ever exhibited such high
bulk ejecta speed. Typically, impostors exhibit much slower outflow
speeds of 300--1000\,km\,s$^{-1}$ (Smith et al.\ 2011).  The
highest-velocity material detected outside $\eta$~Car's Homunculus
nebula only reaches $\sim 5000$\,km\,s$^{-1}$ (Smith 2008).  It is
important to note, however, that the fastest material in $\eta$~Car
traces only a tiny fraction of the outflowing mass (less than 1\%;
Smith 2008), most of which expands at a much slower speed of $\sim
600$\,km\,s$^{-1}$. Emission from the fastest material would not
dominate the shape of Balmer emission-line profiles in the optical
spectrum, as seen during the 2012 outburst of SN~2009ip, even before
the source brightened to normal SN luminosity.

Therefore, the appearance of broad spectral features, particularly the
broad emission, signals the direct {\it transition} of SN~2009ip from
an LBV-like eruption into a true core-collapse SN~IIn. This hypothesis
is further supported by the similarities between the broad spectral
components of SN~2009ip and those of other SNe~II-P, such as SN
2005cs, illustrated in Figure~\ref{fig:spec1}. In this interpretation,
the broad spectral components of SN~2009ip probably represent the
photosphere within the rapidly expanding SN ejecta that is seen
through the CSM, while the narrow emission components could be
generated by interaction of the fastest SN ejecta with the CSM ejected
during the prior eruptions over the past few years, or ionisation
of the CSM by the SN.  SN~2009ip thus joins SN~2006jc as the second
example of a massive star that was actually observed to undergo a
luminous outburst immediately before core collapse.  The fact that
both objects exhibit narrow lines from dense CSM provides strong
support for the conjecture that the CSM of SNe~IIn can be produced by
eruptive precursor events.  SN~2009ip would have significant added
importance, however, since the first outburst was studied in detail,
including high-quality spectra (Smith et al.\ 2010b; Foley et
al.\ 2011), and a luminous blue progenitor star was detected in
archival {\it HST} images (Smith et al.\ 2010b).

Following our initial announcement of high velocities in SN~2009ip
(Smith \& Mauerhan 2012a), two preliminary reports presented optical
photometry which was interpreted as indicating that the most recent
outburst of SN~2009ip was not a true SN (Margutti et al. 2012a; Martin
\& O'Brien 2012).  It was claimed that the peak absolute magnitude of
around $-$14.5 mag ruled out a core-collapse event.  This conclusion
now seems premature, however, since SN~2009ip then rapidly
rebrightened to luminosities consistent with a core-collapse explosion
after all, eventually reaching a peak absolute magnitude of $-$18.  As
shown by our light curve in Figure~\ref{fig:lca}, the initial outburst
reported in July 2012 was evidently a 50--60 day precursor event that
was interrupted by a SN explosion.

\subsection{The Nature of the First 2012 Outburst}

In this section, we focus on the nature of the immediate precursor
outburst in 2012 (before the jump in brightness $\sim 60$ days after
the July 24 rediscovery), and we discuss physical scenarios that could
potentially explain SN~2009ip's behaviour at this early stage.  

While an absolute magnitude of around $-$14.5 mag during this phase
did indeed indicate that SN~2009ip was fainter than {\it normal}
core-collapse SNe, there is a class of faint SNe~II-P with comparably
faint luminosities (Pastorello et al.\ 2007).  The least luminous of
these faint SNe~II-P is SN~1999br, which has a peak of roughly $-$14.2
mag, slightly fainter than the initial 2012 outburst of SN~2009ip.
Relatively low initial luminosities and slow rise times have also been
observed during the early phases other SNe.  Such behaviour was
observed from SN~1987A, and was explained as the result of an
explosion from a blue progenitor with a relatively compact stellar
radius (Arnett 1989, and references therein).  In an explosion of a
compact massive progenitor, the SN might be faint initially due to
adiabatic losses.  Ruling out a core-collapse event altogether for the
precursor outburst of SN~2009ip, based simply on brightness at early
times, is not easily justifiable.

A relatively faint transient might also result from a weak or
``failed'' SN, whereby the collapsing core forms a black hole.  In
such a scenario, the absence of neutrinos leads to a weak explosion
that is unable to completely explode the star, and so most of the
$^{56}$Ni and much of the star's core falls back into the black hole
(Heger et al. 2003). This is expected to produce a transient source
that is fainter than a normal core-collapse event.  Fryer et
al. (2009) predict that such events should have peak magnitudes in the
range $M_V=-13$ to $-15$ mag, comparable to what has been measured for
the first 2012 outburst of SN~2009ip.  With some asymmetry in the
explosion mechanism, this may provide a means to accelerate a small
fraction of the mass in the star's envelope to very high
speeds. However, given the subsequent sharp rise to very high
luminosities shortly after broad lines were seen, this scenario seems
unlikely for SN~2009ip.

A much more likely interpretation is that the 2012 precursor outburst
of SN~2009ip was the result of yet another eruptive phase similar to
its previous outbursts, which in this case occurred immediately before
core collapse. The maximum of $M_V=-14.5$ mag achieved just prior to
the large jump in brightness around Sep. 25 is typical of the peak
luminosities of suspected SN impostors (see Smith et al.\ 2011, and
references therein), and is the same as the previous LBV-like
outbursts of SN~2009ip itself.  This could explain why the SN was
initially so faint, as the mass outflow from the preceding nonterminal
outburst(s) could have partially obscured the SN temporarily.

Indeed, there may be theoretical reasons to suspect a nonterminal
stellar outburst immediately before core collapse. For example, since
the quiescent progenitor was thought to be a very massive star of
50--80\,M$_{\odot}$ (Smith et al. 2010b), it is close to the regime
where the pulsational pair instability could occur (Heger \& Woosley 2002; Heger
et al.\ 2003; Chatzopolous \& Wheeler 2012). The instability could give rise to repeated, 
explosive bouts of oxygen burning, following by eruptive mass ejections during the few years
preceding core collapse.  Smith et al.\ (2010a) noted that the variations of the pulsational pair
instability are essentially indistinguishable (observationally) from
the properties of giant LBV-like eruptions. Alternatively, explosive
nuclear flashes associated with oxygen or silicon burning can
potentially occur near the time of core collapse (Woosley et al. 1973; Smith et al.\ 2011), 
perhaps driven by fluctuations in turbulent kinetic energy (Arnett \& Meakin 2011). Wave-driven 
mass loss associated with those late nuclear burning phases could also be important (Quataert \& Shiode 2012).
 
Whatever the cause of the 2012 precursor outburst during July/August,
the onset of very broad line emission by 2012 Sep. 7--17 signals that
a clear change in the star had occurred \textit{before} the subsequent
jump in brightness near Sep. 25. If SN~2009ip had not yet become a SN
by the time the broad-line emission spectrum appeared, then the event
would be even more difficult to understand, since we would need to
invoke some nonterminal mechanism that is capable of accelerating the
bulk of the stellar envelope to speeds in excess of
8000\,km\,s$^{-1}$, and up to 20,000\,km\,s$^{-1}$ (see
Figure~\ref{fig:spec_he}), far in excess of the escape velocity from
the star's surface -- but without causing the star to brighten yet.
It seems more likely that the appearance time of broad-line emission
marks the initial stages of a terminal SN explosion, to which the
subsequent jump in brightness is causally linked.

\subsection{The Nature of the Second 2012 Outburst}

Beginning near 2012 Sep. 25, SN~2009ip underwent another dramatic
increase in brightness (Brimacombe 2012), rapidly reaching much higher
luminosities, eventually peaking at $\sim 13.6$\,mag in the $R$ band
($M_R \approx -18$ mag; $L \approx 10^{9.1}\, {\rm L}_{\odot}$).
During the rise, the broad Balmer lines in the spectrum diminished,
and they changed into narrower Lorentzian profiles (Smith \& Mauerhan
2012b; Figures 4 and 5). This spectral morphology resembles the
early-time spectra of other luminous SNe~IIn, such as SN~2006gy,
SN~2006tf, SN~1994W, and SN~2011ht, all of which exhibited
broad-winged Lorentzian profiles during their early phases as a result
of electron scattering in the dense SN-CSM interaction zone, while
masking the broad lines from the underlying fast SN ejecta (Smith et
al.\ 2010a, 2008a; Chugai et al. 2004; Mauerhan et al. 2012). Strong
SN-CSM interaction is often accompanied by the formation of an opaque,
cool dense shell (CDS) that forms in the post-shock medium, and this
would explain the almost complete extinguishing of the broad-lined
component in the spectra, as the inner high-velocity ejecta would be
masked by an opaque shell.

The detection of the broad Balmer lines immediately preceding the
rapid increase in brightness on Sep. 25 suggests a causal link between
this event and the precursor outburst, such that the SN had actually
begun \textit{before} the dramatic brightness jump shown in
Figure~\ref{fig:lca}. We therefore favour the interpretation that the
SN explosion had begun by the time we first detected the emergence of
broad-line emission spectrum on 2012 Sep. 7, and perhaps even
earlier. This was well after the onset of the precursor LBV-like
eruption.  In this picture, the subsequent brightness jump on Sep. 25
is dominated by the onset of intense CSM interaction, as the fast
ejecta from an already-occurring SN caught up to the slower material
ejected during previous outbursts. It is interesting to note that the
difference in expansion velocities between the previous LBV-like
outbursts (600--1000\,km\,s$^{-1}$) and the most recent event
(13,000\,km\,s$^{-1}$) would suggest a delay of 1--2 months for the
fastest ejecta to begin overtaking the slower material, which would
imply an approximate explosion date between Aug. 7 and near the time
of discovery in mid-July. As reported by Foley et al. (2012), however,
a spectrum taken on Aug. 26 apparently did not reveal the broad-line
spectrum that we see emerging by Sep. 7 --- but if another LBV-like
outburst was taking place at the time of supernova explosion, as we
have speculated, then it is possible that an optically thick outflow
generated by that event could temporarily mask the high-velocity
ejecta from the emerging SN.

The temporary appearance of very broad He~{\sc i} profiles by Oct. 16
(Figure~\ref{fig:spec_he}), just after peak brightness, indicates a
very dynamic interaction between the SN explosion and CSM. He~{\sc i}
could have been excited as the radiation field from the shocked SN/CSM
interface hardened temporarily, perhaps as a result of the SN ejecta
encountering a slower shell of CSM or a gap between CSM
shells. Indeed, the blue continuum becomes very strong at this stage
(Figure~\ref{fig:spec_he}). The hard UV and X-ray radiation from the
interaction shock would also shine inward, exciting He~{\sc i} in the
fast SN ejecta, thus explaining the broad components of the He~{\sc i}
emission that we temporarily observe on Oct.~16. He~{\sc i} emission
would then fade as the radiation field softened. The evolving
collision between a SN and the CSM, which comprises mass outflows from
multiple (at least three) previous ejections, is potentially quite
complicated, and nonspherical geometry is probably important in the
details of the brightness and spectral evolution.  The small temporary
luminosity increase that occurred after the light curve had begun to
decline (near day 100 in Figure~\ref{fig:lca}) potentially indicates
the onset of interaction with another mass shell or density
enhancement farther out in the CSM.

The nature of the strong double-absorption feature in the range
5600--5800~{\AA} that develops strongly by Nov.~24 is puzzling.
However, the development of strong broad Ca~{\sc ii} around the same
time suggests that Na~{\sc i} is a potential contributor, as these are
both signatures of metal-rich SN ejecta.  The evolution of He~{\sc i}
P-Cygni emission could also have a significant effect on the
morphology of this absorption complex.  In any case, we refrain from
further detailed analysis of this feature at this time. Indeed, the
interaction of SN ejecta with multiple shells of CSM, in adition to a
potentially asymmetric geometrical configuration, will undoubtedly
result in a complicated system of emission components from multiple
zones of shocked CSM and SN ejecta, requiring more detailed analysis.

\subsection{The Nature of the CSM and Prior Outbursts}

Our ability to detect very broad lines just before the main
brightening perhaps indicates that the CSM was partially transparent
at first, compared with some other SNe~IIn such as SN~1998S (Leonard
et al. 2000), SN~2006gy (Smith et al 2007, 2010a; Ofek et al.\ 2007),
SN~2006tf (Smith et al.\ 2008a), SN~1994W (Sollerman et al. 1998; Chugai et
al. 2004), and SN~2011ht (Mauerhan et al. 2012).  All of these SNe had
dense CSM that was opaque to the observational signature of any inner
high-velocity ejecta --- {\it but no SN~IIn has ever been observed in
  the weeks before it brightened}, because in all previous cases the
SN was not discovered and monitored until it had already begun to
brighten substantially.

Thus, we have no precedent for comparison with SN~2009ip.  Perhaps
many SNe~IIn go through a very early phase when their CSM is
transparent, before it is ionised by the CSM interaction and before
the SN brightens.  The early detection of broad lines could simply be
due to the fact that we caught SN~2009ip remarkably close to the time
of core collapse.  This could also be the case if the ejections
associated with the prior outbursts in 2009 and 2010 had a relatively
low mass compared with the outflows of other SN impostors, or if the
CSM had a small geometric covering factor due to a highly clumped CSM
or highly nonspherical geometry (e.g., a disk or bipolar shell). If
so, the progenitor could be an LBV-like star that underwent mass
eruptions similar in scale to those of P-Cygni, which has a nebular
mass of only $\sim 0.1$\,M$_{\odot}$ ejected over 10\,yr (Smith \&
Hartigan 2006). This is in contrast with more massive LBV ejection
nebulae, such as $\eta$~Car's Homunculus or the Pistol nebula, both of
which have masses of $\sim 10$\,M$_{\odot}$ (Smith et al.\ 2003; Figer
et al.\ 1999).

We can use the kinematic information from our spectra combined with
the observed luminosity to constrain the parameters of the CSM.  The
wind-density parameter, which is defined the ratio of the mass-loss
rate ($\dot{M}$) and the CSM velocity ($V_{\textrm{\tiny{CSM}}}$), can
be expressed as
$\dot{M}/V_{\textrm{\tiny{CSM}}}=2~L/V_{\textrm{\tiny{SN}}}^3$, where
$V_{\textrm{\tiny{SN}}}$ is the velocity of the SN ejecta and $L$ is
the luminosity.  Using the peak luminosity $\sim 10^{9}\,{\rm
  L}_{\odot}$, and taking the FWHM of 8000\,km\,s$^{-1}$ of the broad
H$\alpha$ emission line as the average value of
$V_{\textrm{\tiny{SN}}}$, implies a wind-density parameter of $\sim
10^{16}$\,g\,cm$^{-1}$. The absorption-trough minimum of the narrow
H$\alpha$ P-Cygni component from the Nov.~6 spectrum indicates that
$V_{\textrm{\tiny{CSM}}} = -1300$\,km\,s$^{-1}$, implying a
mass-loss ``rate'' on the order of $\dot{M} = 0.01\,{\rm
  M}_{\odot}$\,yr$^{-1}$ for the pre-SN outburst that created the CSM.
These values are to be taken with caution, however. After all, the CSM
around SN~2009ip is likely to be a complicated system, judging by the
persistent and rather chaotic variations and outbursts that have been
observed over the last decade.  Upon close inspection, though, there
does appear to be some evidence for a velocity gradient in the
CSM. For the narrow He~{\sc i} $\lambda$5876 component seen on
Oct. 16, the $-1000$\,km\,s$^{-1}$ velocity of the absorption minimum
in the P-Cygni profile is slightly lower than the value of
$-1300$\,km\,s$^{-1}$ that we measured for the narrow H$\alpha$
component on Nov.~6.  Since the later spectrum traces material further
out in the CSM, the observed velocity gradient may suggest Hubble-like
expansion for the CSM, which we might expect for an LBV mass ejection
(see, e.g., Smith et al.\ 2010a, where this was seen in the case of
SN~2006gy).

There are a few plausible suggestions for the physical nature of the
precursor variability during the past few years.  The most likely
scenario is that the 2009 and 2010 outbursts have the same underlying
cause and mechanism as the 2012 precursor outburst.  As noted above,
possibilities include the pulsational pair instability mechanism or
outbursts associated with late stages of nuclear burning.  However, an
important clue about the cause of SN~2009ip's earlier 2009 and 2010
eruptions might be the short timescale over which they occurred.
Those outbursts lasted for only a few days to weeks, which is distinct
from the much longer lasting outbursts of most other SN impostors (see
Smith et al.\ 2011, and references therein). Such short timescales
suggest a dynamical mass ejection event, rather than a wind-driven
event.  The rapid fluctuations are similar to the variations observed
from the highly variable and recurrent transient SN~2000ch (Pastorello
et al.\ 2010).

The physical reasons for such rapid, large-amplitude variability are
uncertain, but one possibility we might consider is the close
periastron passages of a companion star in an eccentric binary system
(Smith 2011).  This might occur if a companion plunges into the
bloated envelope of an LBV primary, the radius of which may be highly
variable (the change in radius governs whether the violent interaction
occurs). This was suggested by Smith (2011) as a potential explanation
for $\eta$~Car's brief brightening events in 1838 and 1843, both of
which occurred within weeks of periastron encounters (Smith \& Frew
2011).  One would expect such encounters to form a very nonspherical
distribution for the CSM (like a disk or torus), possibly helping to
explain the variable transparency to emission from the inner
high-velocity ejecta in SN~2009ip, as noted above. Whether such
collisional events could help trigger core collapse, however, remains
unknown.

Although the exact causes for the complex pre-SN evolution of
SN~2009ip remain uncertain, the opportunity to spectroscopically
observe such a unique object in the days before it became a SN was
truly remarkable.  Our observations provide dramatic confirmation of
recent conjectures that suggest a link between LBV-like eruptions and
the class of SNe~IIn that are dominated by CSM interaction (Smith \&
Owocki 2006; Smith et al.\ 2010a, 2008a, 2007; Gal-Yam et al.\ 2007;
Gal-Yam \& Leonard 2009). Moreover, the explosion of SN~2009ip
confirms that LBV stars can explode as SNe before becoming Wolf-Rayet
stars, which previously was not expected from stellar models (e.g.,
Langer et al. 1994).  The 2012 event is still underway at the time of
writing, so the details of its late-time evolution should be followed
closely.

\section*{Acknowledgments}

\scriptsize 

We are grateful to the referee, Geoff Clayton, for a careful review of
this manuscript.  We thank Peter Milne for conducting the
Kuiper/MONT4K photometric observations of SN~2009ip, as well as Byung
Yun Choi, Ori Fox, Jenifer Gross, Patrick Kelly, Erin Leonard,
Michelle Mason, and Frank J. D. Serduke for assistance with the Lick
observations. Some of the data presented herein were obtained at the
W.~M.\ Keck Observatory, which is operated as a scientific partnership
among the California Institute of Technology, the University of
California, and NASA; the observatory was made possible by the
generous financial support of the W.~M.\ Keck Foundation.  We are
grateful to the staffs at the Lick (KAIT, Nickel, and Shane
telescopes), Mt.\ Bigelow, MMT, Bok, and Keck Observatories for their
assistance.  This research was also made possible through the use of
the AAVSO Photometric All-Sky Survey (APASS), funded by the Robert
Martin Ayers Sciences Fund.  The supernova research of A.V.F.'s group
at U.C. Berkeley is supported by Gary \& Cynthia Bengier, the Richard
\& Rhoda Goldman Fund, the Christopher R. Redlich Fund, the TABASGO
Foundation, NSF grants AST-0908886 and AST-1211916, and NASA/{\it HST}
grants AR-12126 and AR-12623 from the Space Telescope Science
Institute (which is operated by Associated Universities for Research
in Astronomy, Inc., under NASA contract NAS 5-26555). KAIT and its
ongoing operation were made possible by donations from Sun
Microsystems, Inc., the Hewlett-Packard Company, AutoScope
Corporation, Lick Observatory, the NSF, the University of California,
the Sylvia \& Jim Katzman Foundation and the TABASGO Foundation.  We
dedicate this paper to the memory of Weidong Li, whose unfailing
stewardship of KAIT and enthusiasm for supernova science were
inspirational; we deeply miss his friendship and collaboration, which
were tragically taken away from us much too early.

\end{document}